\documentclass[
    final            
  ]
  {aipproc}
\layoutstyle{6x9}
\newcommand{\NB}{{\sc NBODY4}}

\begin{document}

\title{Gravitational waves from coalescing massive black holes in young dense clusters}


\author{Pau Amaro-Seoane}{
  address={Max Planck Intitut f\"ur Gravitationsphysik (Albert-Einstein-Institut), D-14476 Potsdam, Germany}
}
%
%

\begin{abstract}

HST observations reveal that young massive star clusters form in gas-rich
environments like the Antenn{\ae} galaxy which will merge in collisional
processes to form larger structures. These clusters amalgamate and if some of
these clusters harbour a massive black hole in their centres, they can become a
strong source of gravitational waves when they coalesce. In order to understand
the dynamical processes that are into play in such a scenario, one has to
carefully study the evolution of the merger of two of such young massive star
clusters and more specifically their respective massive black holes. This will
be a promising source of gravitational waves for both, LISA and the proposed
Big Bang Observer (BBO), whose first purpose is to search for an
inflation-generated gravitational waves background in the frequency range of
$10^{-1}-1$ Hz.  We used high-resolution direct summation $N-$body simulations
to study the orbital evolution of two colliding globular clusters with
different initial conditions. Even if the final eccentricity is almost
negligible when entering the bandwidth, it will suffice to provide
us with detailed information about these astrophysical events.

\end{abstract}

\maketitle


\section{Motivation}

Nowadays it is well established that massive stellar clusters form in
interacting galaxies. High-resolution Hubble Space Telescope Observations of
the Antenn{\ae} \citep{WhitmoreEtAl99,ZhangFall99} or Stephan's Quintet
\citep{GallagheretAl01} show that hundreds of young massive star clusters are
lurking in the star forming regions and that they are clustered into larger
clusters of a $\sim$ few 100 pc. The images also reveal that only a few of
these clusters are reddened, thus suggesting that the gas has been already
removed in most of them. Since they harbour $\sim 10^5$ stars within $\sim 1-$
few parsecs and are older than 5 Myr, they are most likely to be bound
clusters. Supernova explosions originating from stars with a mass larger than
$8 M_{\odot}$ only contribute to the total cluster mass -which for $\sim
10^7-10^5$ stars is about ${\cal M}_{\rm cl} \sim 10^7-10^5 M_{\odot}$- in
about $\sim 10\%$.  Stars more massive than $2 M_{\odot}$ contribute in only
$\sim 25\%$ to the total, so that the clusters are very likely to be bound.

The so-called {\em clusters complexes} have been employed in the literature as
a possibility to build compact dwarf galaxies as a result of the amalgamation
of their smaller clusters constituents
\citep{FellhauerKroupa02,FellhauerKroupa05} in collisional processes. On the
other hand, it has been studied by different authors how in such a young dense
cluster collisional processes among the heaviest stars that segregate to the
centre because of dynamical friction might lead to the formation of a very
massive star \citep{PortegiesZwartMcMillan00,GurkanEtAl04,PortegiesZwartEtAl04,
FreitagEtAl06}. Such star will become unstable and thus possibly create a
massive black hole in their centre with a mass ranging between $10^2-10^4
M_{\odot}$ -which therefore receives the surname `intermediate-mass' black hole
(IMBH)-. This means that the young dense clusters in the cluster complexes are
possibly guarding an IMBH in their centres; since the clusters merge with each
other, the respective IMBHs will have, at least, the possibility of forming a
bound system. Whether or not they will merge within a Hubble time and what the
implications for LISA and the BBO would be, is something to be analysed
numerically.

\section{How to merge two young dense clusters}

\begin{figure}
\resizebox{0.5\hsize}{!}{\includegraphics[scale=1,clip]
{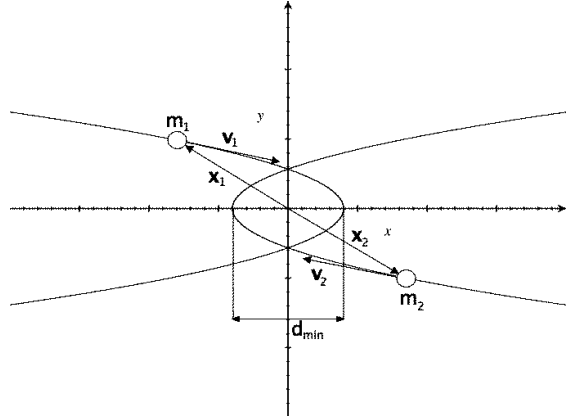}}
\caption{Initial set up for a parabolic collision in the centre-of-mass of
both clusters
\label{fig.parabola}
}
\end{figure}

The clusters were set on to a parabolic orbit so that the minimum distance at
which they pass by is $d_{\rm min}$ of Fig.\,(\ref{fig.parabola}) if they are
considered to be a point particle at that moment. In the centre of mass
reference frame of both clusters, as shown in the figure, we have that ${\bf
x}_1=\lambda_2 \,{\bf d}$, ${\bf x}_2=-\lambda_1 \,{\bf d}$, ${\bf
v}_1=\lambda_2 \,v_{\rm rel}$ and ${\bf v}_2=-\lambda_1 \,{\bf v}_{\rm rel}$;
where ${\bf v}_{\rm rel}$ is the relative velocity of the clusters, 
${\bf x}_{1\,,2}$ their positions (if we regard them to be a point mass, or
to their centres) and $\lambda_{1,\,2}={m_{1,\,2}}/({m_1+m_2})$

From the reduced particle standpoint, we have to determine ${\bf d}$ and ${\bf
v}_{\rm rel}$ when the separation is $d$ (which is given as an initial
condition). Since the reduced mass is $\mu={m_1\,m_2}/({m_1+m_2})$,
for a parabolic orbit we have that the energy at the pericentre

\begin{equation}
E=\frac{-G\,m_1\,m_2}{d}+\frac{1}{2}\mu\, v_{\rm rel}^2\\
=\frac{-G\,m_1\,m_2}{d}+\frac{1}{2}\mu\, v_{\rm max}^2=0.
\label{eq.Epericentre}
\end{equation}

\noindent
Since the specific angular momentum per unit $\mu$ is $l=l_z=-v_{\rm max}\cdot 
d_{\rm min} = |{\bf d}\land {v}_{\rm rel}|_z=x\,v_y-y\,v_x$,
we obtain $v_{\rm max}=\sqrt{{2\,G(m_1+m_2)}/{d_{\rm min}}}$,
for a given specific angular momentum $l=\sqrt{2\,G\,d_{\rm min}(m_1+m_2)}$.

As for the components of the velocities, for a given relative velocity of
$v_{\rm rel}=\sqrt{{2\,G\,d_{\rm min}(m_1+m_2)}/{d}}$ and with the help of the
relations $-l=x\cdot v_y-y\cdot v_x$, $v_{\rm rel}^2=v_x^2 + v_y^2$ and $d^2=x^2+y^2$,
we can infer that the velocity components are

\begin{equation}
v_x=\frac{l\cdot y}{d^2}\Bigg[1+\sqrt{1+\Big(\frac{d}{y}\Big)^2
(x^2\frac{v_{\rm rel}^2}{l^2}-1)} \Bigg]~,
v_y=-\sqrt{v_{\rm rel}^2-v_x^2}.
\label{eq.vel_components}
\end{equation}

\noindent
We resort finally to the definition of parabola to obtain the required
expressions for $x$ and $y$, $2\,d_{\rm min}-x=d$, so that $x=d-2\,d_{\rm min}$
and $y=\sqrt{d^2-x^2}$.

For the model presented in this work, we chose a system with $d_{\rm min}=2$
pc, corresponding to a relative velocity at pericentre of $23.3$ km/s.  The
clusters will always merge and form a larger cluster, because they are
initially set in a parabolic orbit. The number of stars used for each cluster
is ${\cal N}_{\star} =6.3\times 10^{4}$, the masses of they set to $6.3\times
10^{4} M_{\odot}$ and we used for the initial distribution a King model of
concentration $W_0=7$ (cluster 1) and $W_0=6$ (cluster 2).  The central
velocity dispersions were set to $\sigma_{\rm core\,1}=8.41$ km/s and
$\sigma_{\rm core\,2}=8.29$ km/s (hereafter the subscripts 1 and 2 stand for
the cluster 1 and 2), with core radii of $R_{\rm core\,1}=0.203$ pc and $R_{\rm
core\,2}=0.293$ pc. Both clusters host additionally an IMBH of mass $300
M_{\odot}$ in their centres. 

The simulation was performed on a special-purpose hardware GRAPE-6A single PCI
card with a peak performance of 130 Gflops \citep{GRAPE6A}, roughly equivalent
to 100 single PCs, with the direct-summation {\NB} code of Aarseth
\cite{Aarseth03}. This choice was made for the sake of the accuracy of the
study of the orbital parameters evolution of the binary of IMBHs; for this
numerical tool includes both the {\em KS regularisation} and {\em chain
regularisation}, which means that when two or more particles are tightly bound
to each other or the separation among them becomes very small during a
hyperbolic encounter, the system becomes a candidate to be regularised in order
to avoid problematical small individual time steps.  The basis of direct {\NB}
codes relies on an improved Hermit integrator scheme \citep{Aarseth99} for
which we need not only the accelerations but also their time derivative. The
computational effort translates into accuracy and this way we can reliably
follow track of the orbital evolution of every single particle in our system.
Other alternative codes that add a softening to the gravitational forces (i.e.
substituting the $1/r^2$ factor with $1/(r^2+\epsilon^2)$, where $r$ is the
radius and $\epsilon$ the softening parameter) in order to avoid them to become
too large are to be discarded if we want to befittingly make a highly accurate
estimate of the orbital evolution of the IMBHs system (for at a certain point
in the evolution of the binary the separation $\sim \epsilon$) which is the
final purpose of our numerical study.

\section{Dynamics and geometry of the system}

\begin{figure}
\resizebox{0.5\hsize}{!}{\includegraphics[clip]
{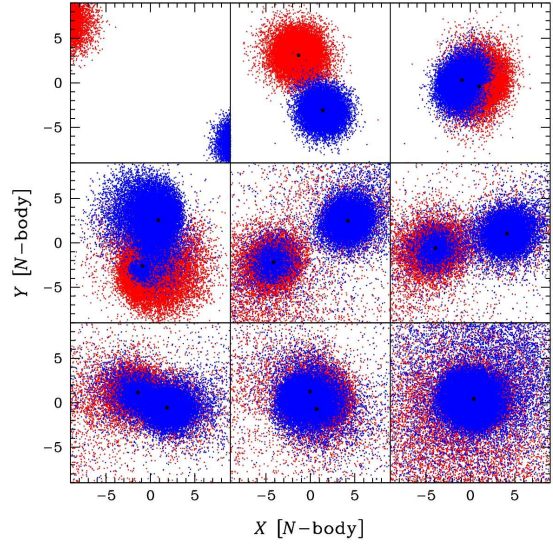}}
\caption{Projection in the X-Y plan of the trajectory of every single particle in
the process of merger of the two young dense clusters. 
See text for further details
\label{fig.mosaic}
}
\end{figure}

\begin{figure}
\resizebox{0.5\hsize}{!}{\includegraphics[scale=1,clip]
{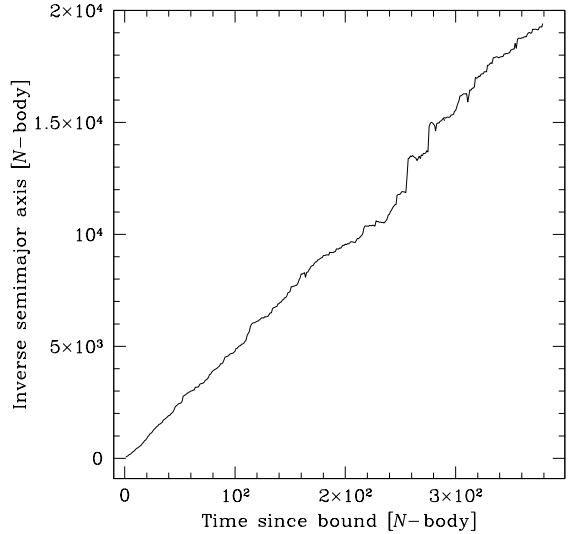}}
\caption{Evolution of the semi-major axis of the massive black hole binary 
\label{fig.InvSemi}
}
\end{figure}

In Fig.\,(\ref{fig.mosaic}) we show  nine snapshots of the merger of the two
clusters and the position of the IMBHs marked with a black dot. The nine
snapshots show the evolution of the clusters merger for $T=0,~ 54,~ 66,~ 76,~
110,~ 128,~ 146,~ 152~{\rm and}~192~ {{\cal U}_{\rm T\, NB}}$ $N-$body time
units, which correspondingly are $T=0, 2.77,~ 3.19,~ 4.62,~ 5.37,~ 6.13,~
6.38~{\rm and}~8.06$ Myrs. We can clearly observe an exchange of stars between
the clusters already after the first cluster interaction (the corresponding
``red'' stars do not show up in the ``blue'' cluster because the blue colour
overwrites the red). As shown in Fig.\,(\ref{fig.InvSemi}), the binary of
intermediate-mass black holes hardens steadily.  We obtain the {\em classical}
value of the {\em hardening constant} $H\simeq 16$
\citep{Quinlan96,SesanaEtAl04}.

\begin{figure}
\resizebox{0.4\hsize}{!}{\includegraphics[scale=1,clip]
{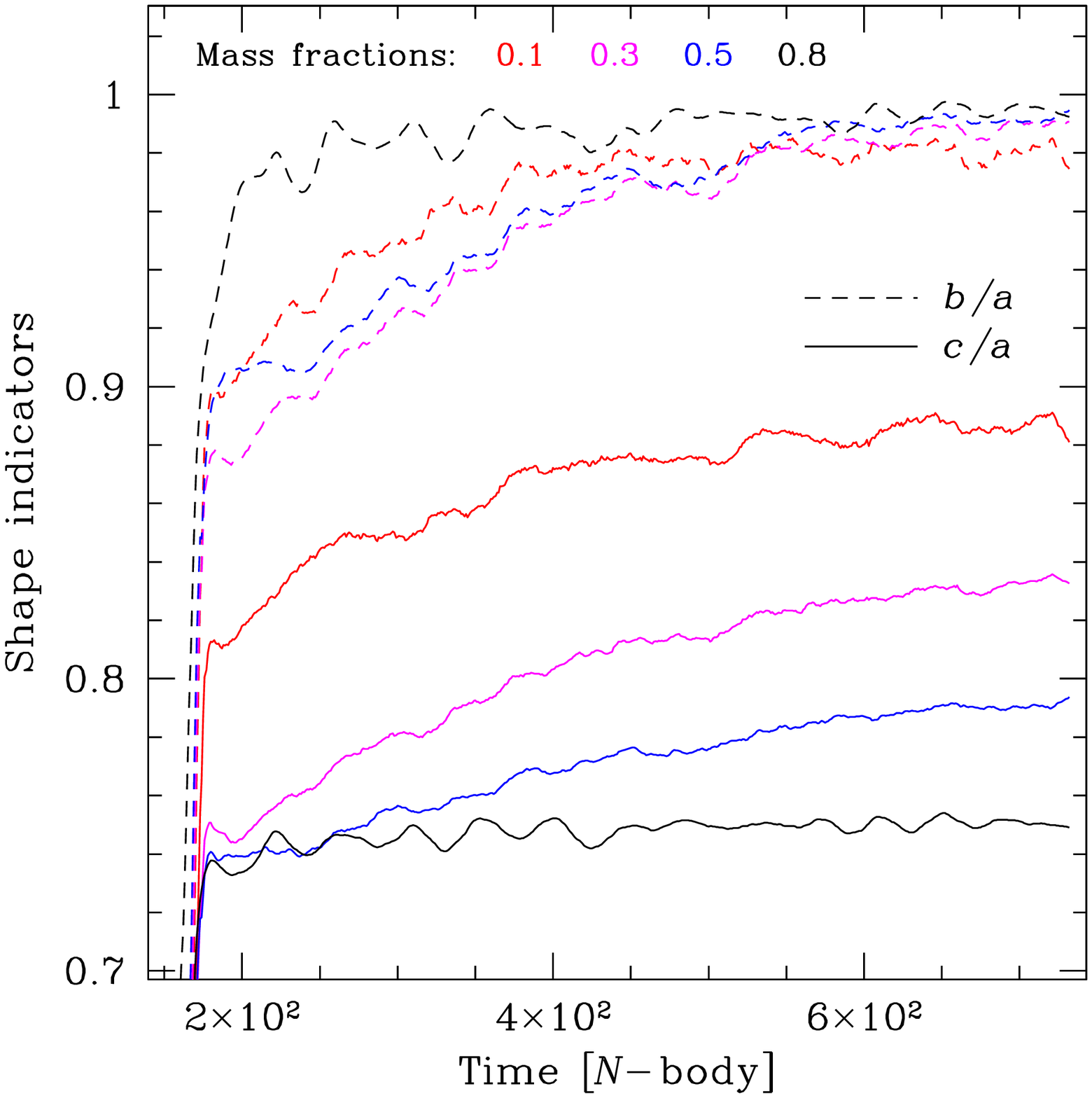}}
\resizebox{0.4\hsize}{!}{\includegraphics[scale=1,clip]
{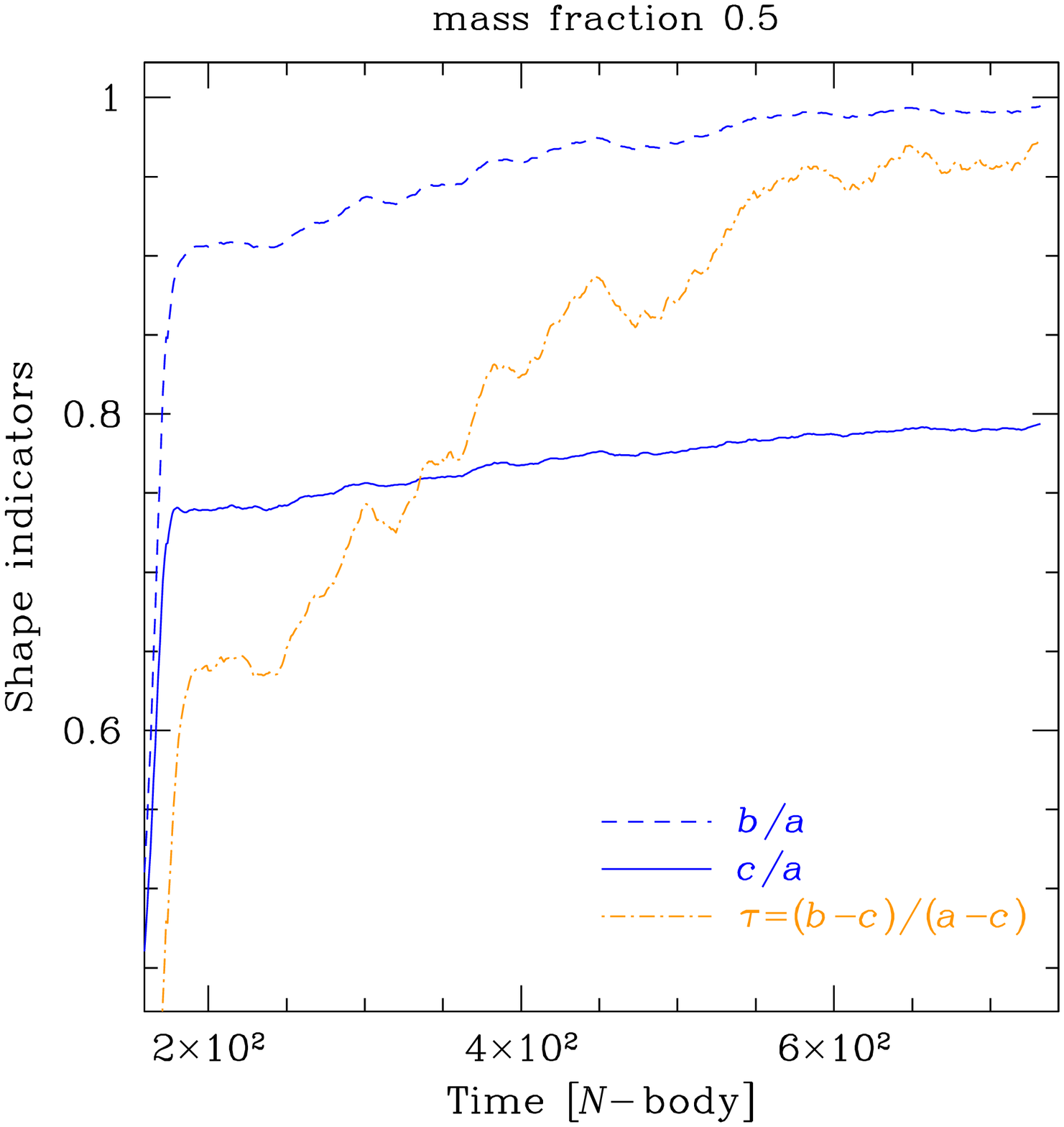}}
\caption{Triaxiality of the resulting merged cluster
for different mass fractions (left panel) and for the
mass fraction 0.5 (right panel)
\label{fig.EvolTriax0p10p8}
}
\end{figure}

In Fig.\,(\ref{fig.EvolTriax0p10p8}) we show the evolution of the triaxiality
of the cluster formed as a result of the merger of the two cluster for our
fiducial model. In the figure, a, b and c --with $a>b>c$ ab definitio-- are the
semi-major axes of the ellipsoid of inertia, determined by four different mass
fractions of the stars which were distributed according to the amount of
gravitational energy. This means that the lower the mass fraction is, the
closer we are to the centre of the resulted merged system.  We can clearly see
that the system resulting of a realistic parabolic cluster merger has a shorter
axis and two approximately equal longer axes, but larger than the third one, $a
\sim b > c$, which means that it is manifestly oblate and not prolate. We can
also observe that the binary of IMBHs makes the system more spherical in the
centre than at larger distances, where it is flatter. This is so because the
flattening depends on the gravitational and centrifugal forces of the cluster,
its size, rotation and density of stars. For the case of two galactic nuclei
harbouring massive black holes, this would translate into a clear danger of
hang-up of the binary or stalling; i.e. there would not be enough stars to
strongly interact with the binary of massive black holes because the loss-cone
-a phase space region where stars can interact with the central object/s in one
crossing time, \citep{FR76,AS01,ASEtAl04}- would remain depleted because of the
lack of centrophilic orbits in an axially symmetric potential. In the case
studied here, though, this is not the case, because the ratio of relaxational
to dynamical timescales is small due to the lower number of stars in the
system.

\section{Implications for LISA and the BBO: General discussion on 
{\NB} prospects for lower-frequency gravitational-waves Astrophysics}

After a strong dynamical evolution with the surrounding stars of the merged
cluster in which the IMBHs are embedded, the evolution is dominated by
gravitational wave emission during the last $10^8$ yrs.  The evolution from the
moment the binary is bound takes $\sim 160$ Myrs and enters the LISA bandwidth
(i.e. has an orbital period of less than $10^4$ sec) on an almost circular
orbit (for a more detailed description of the final stage of the evolution see
\cite{ASF06}). For the work presented here we performed also a parallel test
simulation including the possibility that stars are tidally disrupted by the
IMBHs binary if they enter the tidal disruption radius of one of the IMBHs but
these effects are almost insignificant in terms of the description of the
parameter space of the IMBH binary and thus negligible for the main purpose of
this study. In any case, the ``final'' eccentricity of the binary (the
eccentricity it achieves by entering the LISA band) is almost zero; i.e. the
binary has almost totally circularised.  We can expect an event rate of between
four and five of these events per year for LISA \citep{ASF06}.  As for the
detectability of the gravitational waves emitted by the IMBH binary as it
inspirals and merges, in \citep{ASF06} we show the amplitude ratio assuming a
coherent integration (matched filtering) over the observation time.  For both
observatories, LISA and the BBO, only the n = 2 harmonic of the quadrupolar
emission is clearly detectable during the last few years of inspiral at 1 Gpc,
since it is dominant for low eccentricity binaries; some higher (odd) harmonics
depend on the mass ratio and are zero for equal mass (for zero eccentricity
orbits).  The eccentricity is so low ($10^{-3}$) that the source should be
closer than $\sim 50 - 100$ Mpc to be detectable.

For the data analysis of these sources, it is very important to know the phase
of the gravitational wave; a missmatch even in one single cycle can reduce the
signal to noise ratio (SNR). In practice, since we do not know the parameters
of the gravitational wave signal, we will search over a large parameter space
(including the masses of the objects, the eccentricity etc) and the template
with largest SNR will be chosen. The highest SNR will be achieved by resorting
to templates with parameters close to the actual ones. However, there are some
correlations between the parameters, so that one can compute the level of
correlation and the error bar on the parameter determination.  For strong
signals with long duration within the detector band, we need to know the
waveform (more specifically the phase) very accurately. The eccentricity
affects the phase as well as the amplitude. Since the analysis is very
sensitive to even a tiny change in phase, we can expect that one will be able
to subtract the necessary information of the waveforms of the event to identify
it, due to the induced detectable phase shift of the residual eccentricity
(Bernard Schutz personal communication).

The space-borne LISA mission will fly in $\sim 10$ yrs and critical design
choices affecting the ability to detect this kind of events will have to be
made soon. It is of big importance to produce robust estimates for the rates
and typical orbital parameters of these and other events interesting for the
detection of gravitational waves (extreme mass-ratio inspirals, for instance)
in order to develop a detection template family.  This detection template
family consists of a bank of waveforms. It is relatively fast to generate
waveforms depending on the initial conditions and mass ratio, but not fast
enough if one wants to have a bank of them.  If we want to rigorously explore
the parameter space of these events, we need realistic Astrophysical estimates
of the eccentricity, mass ratio etc at the beginning of the final merger, ``one
step before'' the objects enter the LISA band. An assumption for the initial
parameter space is necessary in order to develop waveform banks. For instance,
in the case of an inspiral search, we know that majority -if not all- of
binaries entering the bandwidth of ground-based detectors like LIGO will be
circular. This allows us to significantly simplify the study of the parameter
space. Similarly, in the case of space-born detectors like LISA, we need to
better understand what we will observe and so focus our efforts on the part of
the parameter space that from which we expect the largest contribution, even if
our final aim is to cover the whole parameter space. For this, an Astrophysical
understanding of the scenario is of paramount importance.  The high-accuracy,
Astrophysical numerical simulations performed with a direct-summation
{\NB} will shed light on many of these aspects by studying realistic
astrophysical scenarios with relativistic corrections, which have been added
recently, and arbitrary geometries \citep{KupiEtAl06,Aarseth03}.
The biggest impact on data analysis is a realistic estimation of the event rate
above a fixed SNR and possibly also as a function of
the SNR itself. The models one can develop with these numerical tools have no
precedents because of the inclusion of non-symmetry, rotation and relativistic
corrections. A realistic estimation of the parameter space is crucial for the
data analysis of LISA gravitational waves. A too small event rate and signals
with smaller SNR will require a method similar to ground based gravitational
waves analysis. This will imply the need for digging out of signals from the
noise with only a few overlapping signals in the data stream.  On the other
hand, in the case of having a large even rate, many overlapped signals would
probably require different data analysis algorithms to make a parameter
estimation.


\begin{theacknowledgments} We acknowledge the Astronomisches Rechen-Institut
for the computing resources on the GRACE cluster of the Volkswagen Foundation,
SFB439. The work of PAS has been supported in the framework of the Third Level
Agreement between the DFG and the IAC (Instituto de Astrof\'\i sica de
Canarias). The author thank the Kavli Institute of Theoretical Physics for
inviting him to attend the Physics of Galactic Nuclei program, where he
finished the manuscript. The work was thus partially supported by the NCF under
Grant PHY99-07949.  He also shows gratitude to Stas Babak, Bernard Schutz and
Curt Cutler for enlightening conversations. The work reported here has been
done in collaboration with Marc Freitag (IoA, Cambridge)

\end{theacknowledgments}



\bibliographystyle{aipproc}   

\bibliography{biblio.bib}

\IfFileExists{\jobname.bbl}{}
 {\typeout{}
  \typeout{******************************************}
  \typeout{** Please run "bibtex \jobname" to optain}
  \typeout{** the bibliography and then re-run LaTeX}
  \typeout{** twice to fix the references!}
  \typeout{******************************************}
  \typeout{}
 }

\end{document}